\documentclass[acmsmall]{acmart}
\setcopyright{none}

\usepackage{enumitem}
\usepackage{caption}
\usepackage{mymacros}
\usepackage{lang}
\usepackage{hyperref}
\usepackage{todonotes}
\usepackage{float}
\usepackage{subcaption}
\usepackage{multirow}
\usepackage{url}
\newcommand{\SOLIDITYKEYWORDS}

\lstnewenvironment{solidity}
    {\lstset{%
      name=main
      language=Java,         
      xrightmargin=2pt,
      xleftmargin=2pt,
      columns=fullflexible,
      keepspaces=true,
      style=numbers,
      classoffset=0,
      frame=bt,
      deletekeywords={byte, char, int, float, double},
      keywordstyle=\color{blue}\bfseries,      
      classoffset=1,
      morekeywords={modifier, mapping, contract, require, uint, emit,
                    assert, if, else, contract, function, returns, var, public}, 
      keywordstyle=\color{brown}\bfseries,  
      classoffset=0,
      commentstyle=\color{blue},             
      morecomment=[l][\color{mauve}]{//},
      morecomment=[l][\color{red}]{//!},
      morecomment=[s][\color{mauve}]{/*}{*/},
      morecomment=[s][\color{red}]{/**}{*/},
      stringstyle=\color{dkgreen},             
      escapeinside={\%*}{*)}                   
}}{}

\lstnewenvironment{solidity-small}
    {\lstset{%
      name=main
      language=Java,         
      columns=fullflexible,
      keepspaces=true,
      style=tinynonumbers,
      classoffset=0,
      frame=bt,
      deletekeywords={byte, char, int, float, double},
      keywordstyle=\color{blue}\bfseries,      
      classoffset=1,
      morekeywords={modifier, mapping, contract, require, uint, emit,
                    assert, if, else, contract, function, returns, var, public}, 
      keywordstyle=\color{brown}\bfseries,  
      classoffset=0,
      commentstyle=\color{blue},             
      morecomment=[l][\color{mauve}]{//},
      morecomment=[l][\color{red}]{//!},
      morecomment=[s][\color{mauve}]{/*}{*/},
      morecomment=[s][\color{red}]{/**}{*/},
      stringstyle=\color{dkgreen},             
      escapeinside={\%*}{*)}                   
}}{}

\AtBeginDocument{%
  }

\newcommand{\ourtechnique}{\emph{Choice-Bound}}

\begin{document}

\title{Universal Scalability in Declarative Program Analysis\\
(with Choice-Based Combination Pruning)}

\author{Anastasios Antoniadis}
\email{anantoni@di.uoa.gr}
\affiliation{%
  \institution{University of Athens}
  \country{Greece}
}

\author{Ilias Tsatiris}
\email{ilias@dedaub.com}
\affiliation{%
  \institution{Dedaub}
  \country{Greece}
}

\author{Neville Grech}
\email{me@nevillegrech.com}
\affiliation{%
  \institution{Dedaub and University of Malta}
  \country{Malta}
}

\author{Yannis Smaragdakis}
\email{yannis@smaragd.org}
\affiliation{%
  \institution{Dedaub and University of Athens}
  \country{Greece}
}

\renewcommand{\shortauthors}{Trovato et al.}

\begin{abstract}
Datalog engines for fixpoint evaluation have brought great benefits
to static program analysis over the past decades. A Datalog
specification of an analysis allows a declarative, easy-to-maintain
specification, without sacrificing performance, and indeed often
achieving significant speedups compared to hand-coded algorithms.

However, these benefits come with a certain loss of control. Datalog
evaluation is bottom-up, meaning that all inferences (from a set of
initial facts) are performed and \emph{all} their conclusions are
outputs of the computation. In practice, virtually every program
analysis expressed in Datalog becomes unscalable for some inputs, due to
the worst-case blowup of computing all results, even
when a partial answer would have been perfectly satisfactory.

In this work, we present a simple, uniform, and elegant
solution to the problem, with stunning practical effectiveness and
application to virtually any Datalog-based analysis. The approach consists of
leveraging the \emph{choice} construct, supported natively in modern
Datalog engines like \souffle{}.  The choice construct allows the
definition of functional dependencies in a relation and has been used
in the past for expressing worklist algorithms. We show a
near-universal construction that allows the choice construct to
flexibly limit evaluation of predicates.  The technique is applicable
to practically any analysis architecture imaginable, since it adaptively
prunes evaluation results when a (programmer-controlled) projection of a
relation exceeds a desired cardinality.

We apply the technique to probably the largest, pre-existing Datalog
analysis frameworks in existence: Doop (for Java bytecode) and the main
client analyses from the Gigahorse framework (for Ethereum smart contracts).
Without needing to understand the existing analysis logic and with
minimal, local-only changes, the performance of each framework
increases dramatically, by over 20x for the hardest inputs, with near-negligible sacrifice in
completeness.
\end{abstract}

\maketitle

\section{Introduction} 

Declarative static program analysis has received significant attention in the past
two decades, with a wealth of research publications and open-source tools~\cite{aplas/WhaleyACL05,DBLP:conf/pldi/MadsenYL16,DBLP:conf/issta/MadsenL18,10.1007/978-3-642-24206-9_14,DataDriven2017,DBLP:conf/pldi/SzaboEB21,madsen13javascript,DataDriven2018,chord,dvanhorn:bravenboer-smaragdakis-oopsla09,Hybrid,ZipperE,Introspective,Scaler,pldi/ZhangMGNY14,pldi/NaikAW06,aplas/LivshitsWL05,pods/LamWLMACU05,DBLP:journals/pacmpl/SzaboBEV18}. The
essence of declarative analysis is to express a program analysis algorithm as
monotonic inference rules that get evaluated up to fixpoint. This style of analysis
specification is an excellent fit for the highly-recursive nature of 
static analysis algorithms, as well as the interdependencies between the analysis
of various distinct program features.

The Datalog language has emerged as the primary platform for declarative program
analysis. Datalog is syntactically similar to Prolog but its evaluation semantics
are strictly declarative. Ordering of rules or of clauses inside an inference
rule does not affect the output of a computation. To maintain this property,
Datalog evaluation (in contrast to Prolog evaluation, which is goal-directed/top-down)
is \emph{bottom-up}: it infers \emph{all} results that follow from the initial input
facts and the transitive application of inference rules, instead of merely attempting
to find one result, or allowing any user control over the search space. In this way,
the Datalog engine is free to decide the evaluation order of rules as well as the exact
implementation of deriving the inference results for a single rule.

Declarative program analysis in Datalog has yielded elegant, concise
analysis specifications that have helped with the invention of new
algorithms (e.g., \emph{type-sensitive} analysis
\cite{popl/SmaragdakisBL11} or \emph{data-driven context tunneling}
\cite{DataDriven2018}).  At the same time, the analysis enjoys great execution efficiency.
For instance, when the \textsc{Doop} framework was introduced~\cite{dvanhorn:bravenboer-smaragdakis-oopsla09}, in 2008,
it outperformed a pre-existing manual implementation of fully equivalent analyses by a factor of 10x.
The reason for this efficiency is primarily that Datalog conducts ``set at a time'' computation,
evaluating rules by joining large tables---a computation that is very efficient
in terms of cache locality, grouping for minimization of overheads, and inherent
parallelism.

At the same time, the not-so-hidden weakness of Datalog-based program analysis has been
its bottom-up evaluation, which starts from input facts and computes \emph{all} possible inferences
from these facts, recursively, until no more inferences can be made. This means that the analysis is very efficient in common cases, but completely
unscalable for pathological inputs (which may have nothing truly ``pathological'', outside the
context of the analysis itself). This occurs most commonly if the analysis input is very large, if the analysis fails to maintain precision
(so that its---final or intermediate---results are large), or if it otherwise explodes in complexity for \emph{part}
of the analyzed program (typically because of a very large number of contexts, in the case of a context-sensitive
analysis). This is a phenomenon commonly identified in the literature---e.g., publication \cite{Introspective} discusses
the abrupt switch to unscalability in length, but virtually any declarative analysis paper mentions unscalability in some form.

Much of the program analysis literature has been about addressing analysis scalability issues,
by producing better algorithms. However, the state of the art remains unscalable for precise context-sensitive analysis of
large inputs (e.g., web applications~\cite{elephant}). Even worse,
the boundary of unscalability can remain entirely unpredictable.

In this work, we introduce \ourtechnique{}: a simple, universal
technique for Datalog analysis scalability, achieving dramatic
results---often outperforming past analysis innovations by
more than an order of magnitude, for the most challenging benchmarks.

There are three significant aspects of the \ourtechnique{} technique: a)
it is simple; b) it is very efficient to implement and leads to great
efficiency/scalability gains; c) it introduces a large design space that enables the
analysis designer to perform intelligent choices.

The main idea of \ourtechnique{} is straightforward. We first observe that Datalog engines can efficiently support functional dependencies:
combinations of certain variables (i.e., columns, when the relation is
viewed as a table) can only occur once. (This is a standard feature
from the data processing days of Datalog, equivalent to declaring
primary keys in database tables.)  One semantics for functional dependencies is
\emph{non-deterministic choice}: keep any one of the value
combinations for the variables/columns that are outside the function domain/key.
The \souffle{} Datalog engine (generally considered to be the most mature, performant, and widely used)
supports this via the \emph{choice-domain} feature~\cite{choice-domain}.

Then, if it is possible to efficiently support non-deterministic functional dependencies
inside a Datalog engine, we can hijack the same mechanism to implement
\emph{multiplicity} dependencies: all that is needed is an extension of the relation
with a shadow variable and a simple hash function. In this way, instead of ``this column combination
can only occur once'' (a functional/key dependency), \ourtechnique{}
expresses ``this column combination can occur at most $N$ times'' (a
multiplicity dependency). This leads to an excellent way to bound and
control analysis complexity, both in terms of internal metrics (e.g.,
``number of contexts'') and in terms of final results (e.g., ``values
per variable''). This adaptation is highly powerful and opens up
a wide design space that the analysis designer can employ for very efficient tuning.

Bounding the number of combinations of key variables
directly addresses the worst-case behavior of the analysis.
\ourtechnique{} effectively solves the problem of scalability of Datalog-based
analyses, at the expense of introducing some analysis incompleteness and
non-determinism.

In overview, the contributions of this work are as follows:
\begin{itemize}
\item We introduce \ourtechnique{}: a way to easily make \emph{any} pre-existing
  Datalog-based program analysis implementation significantly more scalable.
\item We identify the design space behind \ourtechnique{} and introduce a vocabulary
  for expressing the analysis designer's choices. 
\item We show the effectiveness of the technique in major pre-existing analysis
  frameworks and the toughest benchmarks tackled in past literature, achieving
  speedups of over an order of magnitude (and often two or more). We also show
  several metrics of completeness (from coverage to ability to detect bugs)
  and establish that the tradeoff of \ourtechnique{} is excellent in practice.
\end{itemize}

\section{Background}

We begin with some background on program analysis in Datalog and the choice-domain construct.

\subsection{Program Analysis in Datalog}  

Declarative static program analysis (in the Datalog language or its variants, such as Flix~\cite{DBLP:conf/pldi/MadsenYL16,DBLP:conf/issta/MadsenL18} or Inca~\cite{DBLP:conf/pldi/SzaboEB21})
has seen great progress in recent years. The reason is the straightforward specification of highly
complex algorithms and the ability to execute efficiently.

The essence of the Datalog language is its ability to define recursive
relations. Relations (or equivalently \emph{predicates}) are the main
Datalog data type. Computation consists of inferring the contents of
all relations from a set of input relations. For instance, consider the domain
of \emph{pointer analysis}: computing which (object) values flow to which pointer,
throughout the program. If the program-to-analyze is in, say, Java, it is easy to represent the relevant statements
of the Java program as relations, typically stored as database
tables. Consider two such relations, \sv{AssignHeapAllocation(var, obj)} and \sv{Assign(to, from)}. (We follow the convention of
capitalizing the first letter of relation names, while writing
variable names in lower case.)  The former relation represents all
occurrences in the program of an instruction ``\sv{a = new A();}'' (in
Java syntax) where a heap object is allocated and assigned to a
variable. That is, a pre-processing step takes a Java program (most
likely in intermediate, bytecode, form) as input and produces the
relation contents. A static abstraction of the heap object is captured
in variable \sv{obj}---it can be concretely represented as, e.g., a
fully qualified class name and the allocation's bytecode instruction
index.  Similarly, relation \sv{Assign} contains an entry for each
assignment between two program (reference) variables (``\sv{p = q;}'')
in the program. The mapping between the input program and the input
relations is straightforward and purely syntactic. After this step, a
simple, \emph{context-insensitive} pointer analysis can be expressed in Datalog as a
transitive closure computation, as shown in the first five lines of
Figure~\ref{illustration}.

\begin{figure}[htb]
\begin{datalogcode}
VarPointsTo(var, obj) :-
  AssignHeapAllocation(var, obj).

VarPointsTo(to, obj) :- 
  Assign(to, from), 
  VarPointsTo(from, obj).

VarPointsTo(to, obj) :- 
  LoadField(base, fieldname, to), 
  VarPointsTo(base, baseobj), 
  InstanceFieldPointsTo(baseobj, fieldname, obj).

InstanceFieldPointsTo(baseobj, fieldname, obj) :-
  StoreField(base, fieldname, from), 
  VarPointsTo(base, baseobj),  
  VarPointsTo(from, obj).
\end{datalogcode}
\caption{Simple specification of value-flow (points-to) analysis, with \emph{field sensitivity}.}
\label{illustration}
\end{figure}

The Datalog program consists of a series of \emph{rules} that are used
to establish facts about derived relations (such as \sv{VarPointsTo},
which is the points-to relation, i.e., it links every program
variable, \sv{var}, with every heap object abstraction, \sv{obj},
it can point to) based on a conjunction of previously established
facts. We use a Prolog-style left arrow symbol (\sv{:-}) to separate the inferred
fact (the \emph{head}) from the previously established facts (the
\emph{body}). For instance, the second rule in
Figure~\ref{illustration} says that if (for some values of \sv{from},
\sv{to}, and \sv{obj}) \sv{Assign(to,from)} and
\sv{VarPointsTo(obj,from)} are both true, then it can be inferred
that \sv{VarPointsTo(obj,to)} is true.  Note the base case of the
computation (first rule), as well as the (linear) recursion in the definition
of \sv{VarPointsTo} (second rule).

To see the benefits of a declarative specification, consider the rest
of Figure~\ref{illustration}, which refines the rudimentary points-to
analysis of the first two rules, without needing to change them. The
refinement adds to our analysis \emph{field sensitivity}: heap objects
can be stored to and loaded from instance fields and the analysis
keeps track of such actions. (This example ignores other language
features such as method calls---i.e., we assume the analyzed program is
just a single \sv{main} function.) Two new input relations are derived
from the code of a Java program: \sv{LoadField(base,
  fieldname, to)} and \sv{StoreField(from, base,
  fieldname)}. The former tracks a load from the object referenced by
variable \sv{base} in the field identified by \sv{fieldname}.  If,
for instance, the Java program contains a statement ``\sv{x = v.fld;}'',
then \sv{LoadField} contains an entry with the value of
\sv{base} being (a unique identifier of) Java variable ``\sv{v}'',
\sv{fieldname} equal to field ``\sv{fld}'', and \sv{to}
corresponding to ``\sv{x}''.  \sv{StoreField} tracks store
actions in a similar manner: Every Java program statement ``\sv{v.fld =
  u;}'' corresponds to an entry in \sv{StoreField(from,
  base, fieldname)}, with \sv{v} represented by logical variable \sv{base},
\sv{u} represented by \sv{from}, and an identifier for field \sv{fld}
captured by \sv{fieldname}.

The bottom two rules in Figure~\ref{illustration} define and use a new
relation, \sv{InstanceFieldPointsTo}. This computes which heap object
(\sv{baseobj}) can point to which other (\sv{obj}) through a given
field (\sv{fieldname}). The simple definition hides a lot of
conceptual complexity: \sv{InstanceFieldPointsTo} is defined by appeal
to \sv{VarPointsTo} (in two different ways in the same rule), which is,
in turn, defined in terms of \sv{InstanceFieldPointsTo}, and in terms of
itself. The control flow of the analysis is now quite complex
(employing non-linear recursion), but its specification remains
simple.

The striking aspect of the approach is that a) it generalizes
to very complex analyses, both in breadth (i.e., covering all language features) and
in depth (i.e., having a much more precise model); b) it maintains the simple analysis as its core, so that understanding
Figure~\ref{illustration} conveys intuition about analyses with thousands of rules; c) the Datalog code can be
executed highly efficiently.

\subsection{\souffle{}'s Choice Construct} 

The \ourtechnique{} approach leverages the ability to have functional constraints on a relation, with non-deterministic choice
of a representative. This is a feature already implemented highly efficiently in the foremost (in terms of maturity, performance,
and adoption) current Datalog engine, \souffle{}, as the \sv{choice-domain} operator.

Extending Datalog with non-deterministic choice---the ability to choose an aribtrary item from a set---has a long history 
in the database literature\cite{choice-domain, Dusa, NonDet, GreedyChoiceDatalog}. While there have been various mechanisms proposed to achieve this 
goal, \souffle{}'s variant---the choice-domain (or just ``\emph{choice}'') construct---is based on enforcing functional constraints on a relation.
The important features of the construct are, in the words of its authors \cite{choice-domain}, \emph{(1) the simplicity of its semantics, (2) its ease of
implementation, and (3) its efficiency in contrast to having no choice construct
in the language}.

To see how the choice-domain construct works, consider the following program:
\begin{datalogcode}
.decl r(x: number, y: number) choice-domain (x)
.output r

r(1,1).
r(1,2).
r(1,3).
r(2,4).
r(2,5).
r(3,6).
\end{datalogcode}

Let us first focus on the relation declaration: 
\begin{center} \sv{.decl r(x: number, y: number) choice-domain (x)} \end{center}
The start of the declaration is quite standard;
we are declaring a relation \sv{r(x,y)} where both \sv{x} and \sv{y} are numbers. 
We then impose a functional constraint on the relation using the \sv{choice-domain} operator.
Essentially, we are enforcing that for every \sv{x} there can only be one \sv{y}. This is very similar to defining a
primary key in relational databases. With this in mind, and assuming the inferences are processed in the order
of declaration, it is easy to see that the output of the program will be:
\begin{center} \sv{r(1,1) r(2,4) r(3,6)} \end{center}
The rest of the facts get discarded due to the functional constraint. In general, whenever
a new inference is made either directly by a fact or as a result of a rule inference, it is only
added to the relation only if there is no existing tuple in the relation for the columns listed in the choice-domain
declaration.

The choice-domain construct does not add theoretical expressiveness to the language (which is Turing-complete anyway), but
expressing the same constraint in standard Datalog (with stratified aggregation/negation) would have been extremely inefficient
(in addition to very cumbersome and error-prone). It would require computing the unconstrained relation, ordering its
tuples through an arbitrary ordering, and iterating over the ordered tuples to only pick one per functional domain.
In experiments with just a single application of choice-domain to compute a spanning tree, the speedup
is typically over 5 orders of magnitude~\cite{choice-domain}.

Thus, in practice, having language support for non-deterministic choice boosts the expressive power of the language
and allows for more natural and performant implementations of algorithms that rely on a choice mechanism.
The explicit motivation for the construct~\cite{choice-domain} has been worklist algorithms, where picking a representative
(e.g., when forming a spanning tree) is an essential part of the algorithmic logic. Such algorithms arise in
program analysis (e.g., in computing control-flow graphs or strongly-connected components), a domain of prominent
use of Datalog.

The construct can be used for applications far exceeding the original intent, as we shall show. 


\section{\ourtechnique{}} 

We next present the \ourtechnique{} technique at a high-level, discussing its
value proposition.

\subsection{Outline}

The essence of \ourtechnique{} is to lift the choice construct from expressing
\emph{functional dependencies} to expressing \emph{multiplicity dependencies}, and creating
an expressive parameterization space using hash functions. This compact wording
may appear cryptic at first glance, but its essence is simple.

Consider a relation (a.k.a. a \emph{predicate}) \pred{R}{x,y,z}. Applying the choice
construct on \args{x}, \args{y} means that the final contents of \predname{R} will
contain a single \args{z} value for each distinct combination of \args{x}, \args{y}.
Essentially, \predname{R} is declared to be a function from its first two fields
to the last. If, without the choice construct, many combinations of \args{x} and \args{y} were to
appear in the final contents of \predname{R}, there is no guarantee as to which will be kept:
the implementation can make an arbitrary choice, and it is up to the programmer to
ensure that any such choice yields an acceptable output.

The goal of \ourtechnique{} is to relax this constraint from ``a combination of \args{x}, \args{y}
can only appear once'' to ``a combination of \args{x}, \args{y} can only appear up to $N$ times.''
Keeping with the spirit of the choice construct, if more than $N$ combinations were to
arise (without the choice construct), there is no guarantee as to which $N$ combinations will be kept.
The choice will be arbitrary, based on the evaluation order that the implementation chooses.

This relaxation is achieved by adding an extra field to the relation that is choice-bound, denoting
a unique identifier of the ``choice'' that needs to be made. If these identifiers are then
computed from a bounded domain of size $N$, we have the desired guarantee. Specifically, in our
example:
\begin{itemize}
  \item relation \pred{R}{x,y,z} becomes \pred{R}{x,y,z,i} with choice-domain (in the standard
    ``functional'' sense) \args{x,y,i}.
  \item Field \args{i} can then be computed via a hash function
    over the values of \args{z}, projected modulo $N$.
  \item In this way, \pred{R}{x,y,z,i} is computed to have a single instance of each unique combination
    of \args{x,y,i}, i.e., up to $N$ different \args{z} values for each combination of \args{x,y} values.
\end{itemize}

This seemingly simple approach yields tremendous power (discussed
extensively in Section~\ref{sec:design-space}), by appropriately
selecting the choice-bound fields, the bound parameter $N$ itself, but
also the domain of the hash function.


\subsection{Discussion of Impact}
\label{sec:discussion}

Before we delve into the technical specifics of \ourtechnique{}, it is worth understanding
its cost-benefit proposition.

In substance, \ourtechnique{} enables an analysis designer to:
\begin{itemize}
\item express their code declaratively, as before, with virtually no changes to the rules;
\item tune performance by adding extra constraints of the form ``relation \predname{R} shall never
  have more than $N$ instances with the same values for fields \args{x,y}''. If more
  results are going to be produced during rule evaluation, an arbitrary choice of which
  results are kept (typically the ones derived earlier) is made.
\end{itemize}

Thus, the value of \ourtechnique{} is that it can bound the cost of
computation, prevent worst-case blowup of analysis time, as well as
prune analysis paths that are unlikely to be fruitful (e.g., cover the
same code element or combination of conditions, when they have already
been covered $N$ times). 

If the underlying analysis already scales well, \ourtechnique{} has little to offer.
However, even the most well-tuned, practical analyses can easily become unscalable, for a variety
of reasons. These reasons can include larger inputs~\cite{elephant}, or
corner cases where precision is not preserved, leading to unscalability. (Even the simplest points-to analysis,
for instance, is fundamentally an $O(n^3)$ computation~\cite{sas/SridharanF09}. It is not
rare for a small part of the program to trigger such worst-case behavior, even leaving the rest
of the program's analysis unaffected. However, if left unbounded, this worst-case behavior can
easily dominate analysis time.)

In practice, bounding computation costs allows \ourtechnique{}
to gain orders-of-magnitude increases in efficiency and scalability.
The apparent trade-off is to sacrifice completeness and determinism.

This is an excellent trade-off in practice: most
realistic analyses we have found have already made design decisions to
sacrifice completeness and determinism, in order to get much more
minor benefits than those offered by \ourtechnique{}. What
\ourtechnique{} does is empower the analysis designer to find the
sweet spot in this trade-off.

The scalability problem with most large-scale Datalog-based analyses stems from their bottom-up
nature: the analysis computes \emph{all} results, following from the inference rules. However,
in practice it is commonly the case that not all results are necessary and that \emph{which
specific} results end up being computed matters little.

Consider a points-to analysis, such as the ones supported in the \textsc{Doop} framework---probably
the largest single artifact of Datalog analysis code in existence, with over 45KLoC of Datalog code (in over 5,000 rules).
The analyses produce
points-to/value-flow inferences, together with a call-graph. The main final results of the
analysis (consumable by other clients, human eyes or programmatic) are:
\begin{itemize}
\item The points-to information itself, i.e., the values that a given variable can take.
\item The call-graph/reachability information for program functions.
\item Information that results in bug warnings, such as taint information~\cite{ptaint}.
\end{itemize}

All of these outputs already tolerate incompleteness and non-determinism.
Notably, whole-program static analysis (which is the kind that can become unscalable)
does not typically establish the \emph{absence} of some values from a value set, only estimates
the \emph{presence} of values. For instance, although the analysis may compute virtual calls
with a unique target and casts that can never fail, this is done as a metric and not because
this information is actually actionable for optimization: complex language features, such as
reflection and dynamic loading~\cite{soundiness}, make this information incomplete, i.e., unreliable
for automatic application. In general, missing
some points-to/value-flow analysis results is an inevitable fact of life that
analysis designers are already comfortable with.

Similarly, non-determinism is expected. The \textsc{Doop} output is non-deterministic, due
to multiple factors, such as multi-threaded fact generation, multi-threaded core analysis,
arbitrary choice of representatives (e.g., for string objects or reflection~\cite{coloring}),
and much more. To quantify just one of these sources of non-determinism: a quick text search over the \textsc{Doop} code reveals over 680 uses
of the \souffle{} Datalog ``\sv{ord}'' operator, which returns an ``ordinal'', i.e., a unique identifier
(typically an implementation-internal pointer address or index) for an entity. Each such use is a point
of non-determinism in the analysis
logic: re-running the analysis can make \sv{ord} return a different value. The
operator is typically employed in order to come up with \emph{some} ordering of entities (e.g.,
to form spanning trees by selecting a representative from an equivalence class,
or to exhaustively iterate over all entities of a certain kind).

Therefore, in principle, \ourtechnique{} sacrifices nothing that analysis designers
have not alreadyd decided to sacrifice. The question remaining is one of the \emph{extent} of the
sacrifice, which is an engineering trade-off question, answered experimentally. As we shall see in
Section~\ref{sec:evaluation}, analysis incompleteness is very minor, whereas the performance
benefits are dramatic. This is also largely due to the expressive richness of
\ourtechnique{}, discussed next.

\section{\ourtechnique{} Design Space and Applicability}
\ourtechnique{} admits a large design space for parameterizing an analysis. We discuss
the options, as well as practical implementation,
illustrated with realistic examples.

\subsection{Design Space and Vocabulary}
\label{sec:design-space}

\ourtechnique{} opens up a large design space for parameterization of a Datalog computation.
This design space permits bounding, e.g., the internal complexity of an analysis, the final
observed values, key intermediate concepts, etc. This expressiveness permits precise tuning of
a bounded computation.

To encode the design space available to the programmer, we introduce a vocabulary for describing
choice-domain decisions under the \ourtechnique{} technique. The vocabulary admits three
parameters:

\begin{itemize}
\item The \emph{bound} variables, i.e., the dimensions/fields of a predicate whose combinations
  will be limited;
\item the \emph{limit}, i.e., the numeric bound;
\item the \emph{counting} variables, i.e., the dimensions/fields of
  the predicate that participate in the hash function, getting mapped
  to a unique identifier. (In all examples so far, we have set the
  counting fields to be all the fields that are not bound, but this is not necessary,
  as we shall soon discuss.)
\end{itemize}

We use the notation \choicedomain{Relation}{bound vars}{limit}{counting vars} to capture this design
space.

For illustration, let us consider predicate \predname{VarPointsTo}, the main relation of a context-sensitive
points-to analysis. This relation appears in numerous practical frameworks and research
publications~\cite{dvanhorn:bravenboer-smaragdakis-oopsla09,10.1007/978-3-642-24206-9_14,Hybrid,ZipperE,Introspective,Scaler}
and forms the core concept behind most published pointer analysis
algorithms.

The relation has 4 dimensions:
\pred{VarPointsTo}{var,ctx,hobj,hctx}. The intuitive meaning is that
local variable \args{var}, under context \args{ctx}, may point to
abstract \emph{heap object} \args{hobj}, which was originally created under
\emph{heap allocation context} \args{hctx}. (The exact definition of contexts is orthogonal
to our present discussion. It is only worth noting that it can vary
tremendously, to yield different analysis algorithms.)

We consider several design choices over the same relation (and even the same underlying
hash function). All of the design choices yield different performance/completeness profiles for the exact
same core analysis. We illustrate, non-exhaustively, some of the options for predicate \predname{VarPointsTo},
together with an intuitive explanation of the design intent behind each option.
For concreteness, we use specific numbers for the \emph{limit} parameter,
but changing this limit is an obvious parameterization choice, which can be guided by observation,
so we do not discuss it further in conceptual terms.

\begin{itemize}
\item \choicedomain{VarPointsTo}{var}{541}{ctx,hobj,hctx}: This is a straightforward ``cost-conscious'' design
  choice. It limits the total output tuples/entries per local variable to at most $541$ and every combination
  of context, heap object, and heap allocation context counts as a different tuple against this limit.
  Effectively, such a design choice bounds the \emph{cost} of the analysis and not its externally
  observable outputs: each variable can contribute at most $541$ entries to the final relation, without
  consideration regarding the meaning of these entries. For instance, all entries could have the variable
  possibly pointing to the same abstract value.
  
\item \choicedomain{VarPointsTo}{var,ctx}{101}{hobj,hctx}: This is a balanced design choice that mixes
  externally observable quantities (variables and heap objects) with internal elements that add precision
  at the expense of extra computation cost (contexts). The design choice limits every context-qualified variable
  to appearing at most $101$ times in the analysis output. These $101$ occurrences are identified
  by hashing both the heap object and the heap allocation context. It is perfectly possible for
  all $101$ combinations of a context-qualified variable to point to the same heap \emph{object} \args{hobj}
  but with different heap allocation contexts, \args{hctx}. This is notably different from the next
  design option.

\item \choicedomain{VarPointsTo}{var,ctx}{67}{hobj}: This option maintains the spirit of the preceding one,
  but with a restrictive twist. Only the heap \emph{object} participates in the unique identifier of the
  up-to-$67$ allowed combinations for each context-qualified variable (\args{var,ctx}). This means that if the
  same heap object under two different heap allocation contexts, \args{hctx}, is computed to reach the same
  context-qualified local variable, only one of the two tuples will be kept. In effect, heap allocation
  contexts are treated as less valuable information than heap objects: we can have many heap objects appear
  for a context-qualified variable, but with only one heap allocation context each.

\item \choicedomain{VarPointsTo}{var,ctx,hctx}{31}{hobj}: This design option captures yet another
  tradeoff, where the limit is in the number of values finally computed under the full precision of
  the analysis, i.e., as much context information as can be kept. One can see such a choice as
  saying ``if the analysis computes too many heap objects, even with the full precision of
  variable contexts and heap allocation contexts, then it is not fruitful to keep all of them---just
  keep $31$''.
  
\item More choices are possible. These include \choicedomain{VarPointsTo}{var,ctx,hobj}{13}{hctx}
  (where the bound is on the number of heap allocation contexts kept, for all other dimensions being
  identical), \choicedomain{VarPointsTo}{var,hobj}{277}{ctx,hctx} (where there is a total limit
  in the number of appearances of a variable-value pair, which is an externally-observable quantity); etc.
\end{itemize}

\subsection{Applications}
\label{sec:applications}

We applied \ourtechnique{} to several pre-existing Datalog analyses (with minimal change to the analysis
specification, as discussed in Section~\ref{sec:implementation}). We discuss next the design choices
for each of these frameworks, using our vocabulary of the previous section.

\subsubsection{\textsc{Doop}}
The \textsc{Doop} framework for points-to and taint analysis of Java+Android bytecode
is probably the largest Datalog analysis codebase. As already mentioned, it comprises over 5,000 rules, in over 45KLoC.
Applying \ourtechnique{} consisted of bounding only the \predname{VarPointsTo} relation. (\pred{VarPointsTo}{var,ctx,hobj,hctx} is the main
points-to relation, discussed in Section~\ref{sec:design-space}.)

The bound used for the 2-object-sensitive+heap (2objH) analysis of \textsc{Doop} (i.e., an object-sensitive~\cite{dvanhorn:milanova-etal-tosem05}
analysis with 2 elements of context for local variables and 1 for heap objects) is:
\begin{itemize}
\item
    \choicedomain{VarPointsTo}{var,ctx}{101}{hobj,hctx}.\\
    That is, each context-qualified variable is limited to pointing to 101 context-qualified abstract objects.
    This is the ``balanced'' choice discussed earlier.
    
\end{itemize}

Good results can be obtained with various other design choices. However, the above settings are indicative
of the potential of the approach.

Notably, \textsc{Doop} is a framework that encompasses many tens of analysis algorithms, for different flavors of context sensitivity.
The \ourtechnique{} bound will likely need to be adapted for different kinds of context sensitivity, since, for instance, it makes
no sense to use a bound of $101$ context-qualified objects for an analysis with shallower (or none), or different-flavor context.
However, there are not many analyses that are heavy or useful enough to be worth defining bounds for (or if they do become useful
enough, experimenting with an appropriate bound is minor overhead). The 2objH analysis is by far the most prominent analysis:
it is highly-precise for real applications but has failed in the past to scale to anything but small ones.

\subsubsection{Symbolic Value-Flow Analysis}

Symbolic value-flow (\emph{symvalic}) analysis~\cite{symvalic} is a
recent large-scale analysis framework for Ethereum VM smart contracts,
built on top of the Gigahorse decompiler~\cite{gigahorse}. The symvalic analysis core and its clients
comprise some-3,000 Datalog rules, over nearly 30KLoC.
We applied \ourtechnique{} to two relations:
\begin{itemize}
\item \pred{ExprFlowsToVar}{complexity,expr,ctx,var} is the relation
  responsible for the symvalic pre-analysis~\cite[Sec.4.2]{symvalic},
  which creates a universe of symbolic expressions. The meaning of the relation is
  ``variable \args{var} under context \args{ctx} can refer to expression \args{expr}, which has the
  listed \args{complexity}.'' The complexity is used for only allowing a limited number of
  processing steps while defining the universe of expressions, since this analysis can create an unbounded
  number of expressions (e.g., all integers) even for a very small expression size.

  The \ourtechnique{} specification for this relation is:\\
  \choicedomain{ExprFlowsToVar}{var}{1373}{complexity,expr,ctx}.\\
  Thus, in this case we have a pretty large bound but over all possible tuples that pertain to a single
  variable. Intuitively, the reason that this design choice is different from others is that the
  relation being bounded does not keep any real precision: it computes \emph{all} possible expressions
  in the universe, up to a finite number of combinations of program operations. Therefore, it makes sense
  to merely bound the total weight of a single variable, i.e., to have a more cost-conscious bound, as opposed
  to a balanced one.

\item \pred{VarMayBe}{var,expr,deps} is the main relation of the symbolic value-flow analysis, indicating
  that a variable \args{var} may have as its value a concrete or symbolic expression, \args{expr}, and for
  this to happen several dependencies, \args{deps}, also need to hold. (The shape of ``dependencies'' is
  orthogonal to our discussion. Briefly, they are of the form
  ``argument $a$ of the enclosing function needs to have value $v$'', ``global variable $g$ needs to have
  value $o$'', etc.~\cite{symvalic}.)
  
  The \ourtechnique{} specification for this relation is:\\
  \choicedomain{VarMayBe}{var,deps}{257}{expr}. \\
  That is, the bound is straightforward, limiting the possible values of a variable, under the full
  precision of the analysis (i.e., without limiting the dependencies kept) to at most 257.

  Note how these bounds are higher than the earlier bounds in the \textsc{Doop} framework analyses.
  This is easy to understand by considering that the \textsc{Doop} framework is meant for whole-program
  analysis of applications with many tens of thousands of classes (and many tens of MB in binary form),
  whereas symvalic analysis attempts a much more precise analysis but over programs (smart contracts)
  that are at most 24KB in binary form.
\end{itemize}

\subsubsection{Other applications}
We have additionally applied \ourtechnique{} to other analyses, not as
prominent as the above, nor as mature and independently developed. For
instance, one of the applications is to a symbolic execution engine,
where the Datalog analysis leads to queries to an SMT engine. An
excellent bound in this case is to the number of symbolic conditions that ``cover'' the same
program block. This leads to bounding the number of SMT invocations and not just Datalog analysis, yielding
very substantial speedups.

The overall message is that the technique applies to virtually any demanding Datalog analysis and its
expressiveness permits very powerful tuning in order to find sweet spots in the
performance-completeness trade-off.

\subsection{Implementation}
\label{sec:implementation}

Implementing \ourtechnique{} is simple, requiring only local changes to a Datalog program
(modulo a global semi-automatic search-and-replace). As a result, we are able to apply \ourtechnique{}
to virtually \emph{any} Datalog program analysis, with limited understanding of its logic
or internals, as long as we can reason about the parameter space in the terms defined in Section~\ref{sec:design-space}.

The specific implementation technique that we use consists simply of the following steps.
\begin{itemize}
\item Identifying the relation to bound and deciding on the choice of parameters, per Section~\ref{sec:design-space}.
  As a running example, consider relation \predname{VarPointsTo} of the \textsc{Doop} framework, bounded as
  \choicedomain{VarPointsTo}{var,ctx}{101}{hobj,hctx}.
\item Creating a ``bounded'' version of the relation, in addition to the original:
\begin{datalogcode}
.decl VarPointsTo_Bounded(var:Var, ctx: Context,
                          hobj: Value, hctx: HContext, n: number)
       choice-domain (ctx, var, n)
\end{datalogcode}
\item Having the original relation obtain its contents from the bounded one:
  \begin{datalogcode}
VarPointsTo(var,ctx,hobj,hctx) :-
   VarPointsTo_Bounded(var,ctx,hobj,hctx,_).
  \end{datalogcode}
\item Defining the hashing that will produce ids for combinations of the \emph{counting} variables, via a macro:
  \begin{datalogcode}
#define BOUND_VAR_POINTS_TO(var,ctx,hobj,hctx) \
   VarPointsTo_Bounded(var,ctx,hobj,hctx, (ord(hobj)*ord(hctx)) 
  \end{datalogcode}
  The macro uses the \souffle{} \sv{ord} operator to obtain an (arbitrary) internal identifier per object.
\item Performing a global search-and-replace to substitute \sv{BOUND\_VAR\_POINTS\_TO} in the heads of rules
  that would otherwise produce \predname{VarPointsTo} results. The \emph{uses} of the original relation (i.e.,
  in the bodies of rules and not in the head) remain unaffected.

  Furthermore, not even all rules with the bounded
  relation in their head need to be affected: e.g., tedious rules that produce few results (e.g., initialization logic)
  can remain in their original form, since there is little need to limit their inferences. If we want the results of
  such rules to also participate in the bounding (so that we have a guarantee for the total tuples, as per the bounding
  policy), we can also include a feed-back rule:
\begin{datalogcode}
BOUND_VAR_POINTS_TO(var,ctx,hobj,hctx) :-
   VarPointsTo(var,ctx,hobj,hctx).
\end{datalogcode}
  In this way, if tuples of the relation can be derived by non-bounded rules, they still participate in the bounding
  policy. 
\end{itemize}

As an end result, \ourtechnique{} can be used to obtain dramatic scalability benefits within mere hours of
adaptation work, agnostically, for nearly any Datalog-based program analysis.

\section{Evaluation} 
\label{sec:evaluation}

We applied \ourtechnique{} to large pre-existing Datalog static analysis frameworks, as discussed in Section~\ref{sec:applications}.
We next evaluate the performance and completeness, relative to the unmodified code.

\subsection{\textsc{Doop}}
We evaluate \ourtechnique{} in the \textsc{Doop} framework, executed with the \souffle{} Datalog engine. We perform the evaluation on a dataset that consists of all programs we could find in past literature for which \textsc{Doop} had trouble scaling. Most of these programs
are well-known open-source web applications~\cite{elephant}, whereas some more are from the DaCapo benchmark sets~\cite{dacapo:paper}.

We evaluate with the default 2-object sensitive analysis configuration in \textsc{Doop} using the Java 11 library. This is the \textsc{Doop} analysis that one would always
\emph{want to} run for precision, but generally cannot because it fails to scale to large programs.

We use a machine with two Intel(R) Xeon(R) Gold 6136 CPUs @ 3.00GHz (each with 12 cores x 2 hardware threads) and 640GB of RAM. Each benchmark is analyzed separately, with 12 threads, and we evaluate on two dimensions: completeness and speed.

Our benchmarks include:

\begin{itemize}

\item \emph{alfresco}: An open-source content management system (CMS) and business process management software. It helps organizations manage documents, collaboration, records, and content online, enabling digital workflows and secure document handling. (GitHub, 142 stars, 82 forks.) Application classes: 9164. Total classes: 37163.
\item \emph{batik}: An open-source software toolkit for handling Scalable Vector Graphics (SVG) in Java that provides a suite of tools for viewing, generating, and manipulating SVG graphics. Developed by the Apache Software Foundation. (DaCapo benchmark) (GitHub, 214 stars, 143 forks.) Application classes: 2511. Total classes: 11993.
\item \emph{bitbucket-server}: A self-hosted Git repository management tool designed for collaborative software development. Bitbucket Server allows teams to host their Git repositories on their own servers (rather than in the cloud) and provides features like code review, branch permissions, and integration with other tools. Developed by Atlassian. Application classes: 581. Total classes: 29984.
\item \emph{bloat}: Despite being a mere software engineering project by a Ph.D. student, bloat is notorious for its complexity and the scalability challenges it poses. (DaCapo benchmark) Application classes: 360. Total classes: 4813.
\item \emph{dotCMS}: An open-source, hybrid content management system (CMS) designed for businesses that require a flexible and scalable platform to manage content across multiple channels. (GitHub, 864 stars, 468 forks.) Application classes: 5473. Total classes: 46027.
\item \emph{opencms}: An open-source content management system (CMS) offering a powerful solution for creating and managing websites, intranets, and online applications. Developed by Alkacon Software. (GitHub, 528 stars, 575 forks.) Application classes: 2143. Total classes: 16183.
\item \emph{pybbs}: PyBBS (Python Bulletin Board System) is an open-source, web-based bulletin board software developed in Python. PyBBS provides a platform for online discussions, allowing users to post messages, reply to threads, and communicate in a community setting. (GitHub, 910 stars, 487 forks.) Application classes: 172. Total classes: 24692.
\item \emph{shopizer}: Shopizer is an open-source Java-based e-commerce software designed to help businesses build online stores and manage product catalogs, orders, customers, and other e-commerce operations. (GitHub, 3,600 stars, 3,000 forks.) Application classes: 1151. Total classes: 35484.
\item \emph{jython}: An implementation of the Python programming language that runs on the Java platform. Essentially, Jython allows Python code to be executed within the Java environment, providing compatibility and interoperability between Java and Python. (DaCapo benchmark) (GitHub, 1,200 stars, 192 forks.) Application classes: 919. Total classes: 7303. 
\end{itemize}

\subsubsection{Performance}

\begin{figure}[tb!]
  \begin{tabular}{lrrr}\toprule
  \textbf{Benchmark} &\textbf{2obj Execution Time} &\textbf{\ourtechnique{} Execution Time} \\\midrule
  alfresco &86400 (Timeout) &1,640 \\
  batik &1,287 &274 \\
  bitbucket-server &9,824 &560 \\
  bloat &1,269 &80 \\
  dotCMS &86400 (Timeout) &3,467 \\
  jython &86400 (Timeout) &1,706 \\
  opencms &10,549 &593 \\
  pybbs &8,260 &568 \\
  shopizer &8,868 &509 \\
  \bottomrule
  \end{tabular}
\caption[]{Execution times shown in seconds for the default 2-obj analysis and \ourtechnique{}.}
\label{fig:executiontimes}
\end{figure}

\begin{figure}[tb!]
  \includegraphics[scale=0.6, clip=true, trim=20 10 2 10]{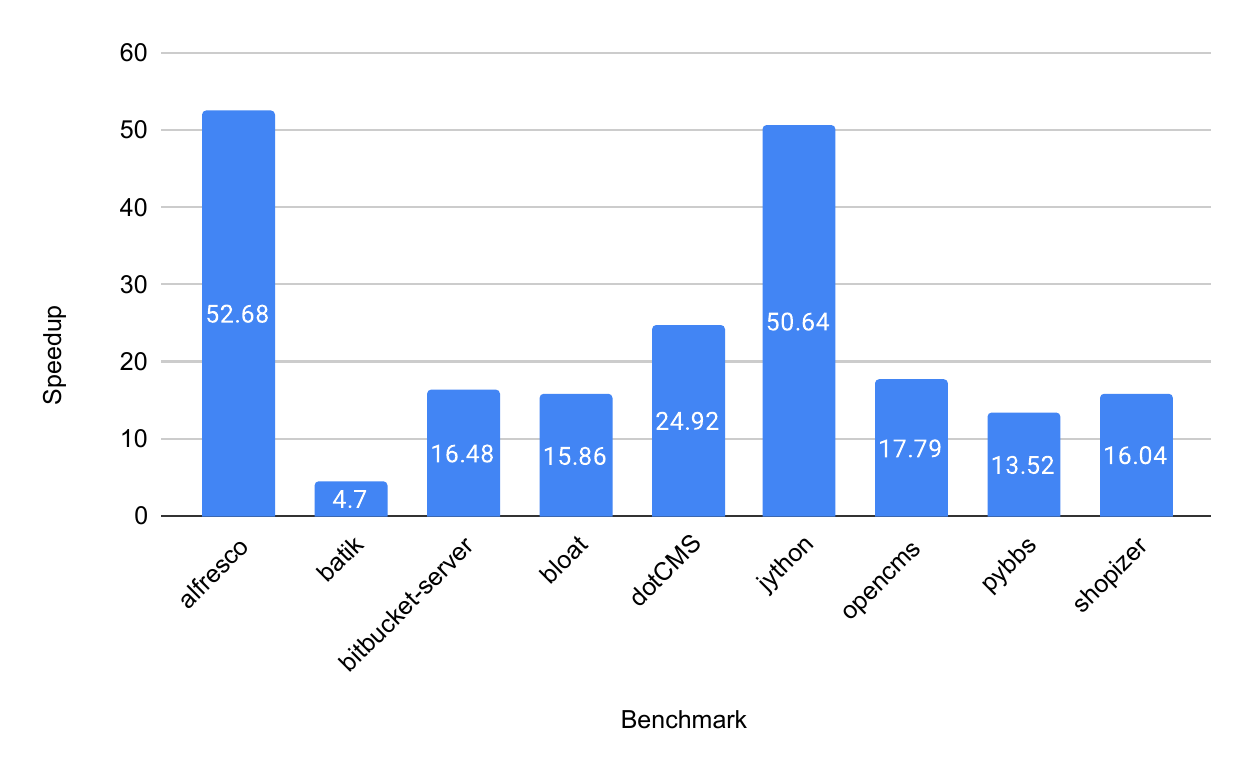}
  \caption{Speedup per benchmark. The numbers shown are \emph{multiplicative factors}. The three highest bars are under-estimates, since the default analysis never
  terminated, in 24hrs.}
\label{fig:speedup}
\end{figure}

The most important aspect of \ourtechnique{} implementation is its impact on analysis performance. All 9 benchmarks terminated in under 1 hour and typically in under 30 minutes. This includes benchmarks that do not terminate in multiple days with the default analysis.

Figure~\ref{fig:executiontimes} tabulates the execution times of the original analysis and the \ourtechnique{} analyis over the benchmark set. Figure~\ref{fig:speedup} plots
the speedup.

As can be seen, the speedup is dramatic.  In the case of
\emph{alfresco}, \emph{dotCMS}, and \emph{jython}, if we consider the
24-hour timeout limit as a lower bound for their execution time, we
see a >52.68x speedup for \emph{alfresco}, a >24.92x speedup for
\emph{dotCMS} and a >50.64x speedup for \emph{jython}. In more
realistic terms, however, a terminating analysis is arguably
immeasurably better than an analysis that times out. It is also worth
noting that even a timeout of 48 hours is not enough for a 2-object
sensitive analysis to terminate for either \emph{alfresco},
\emph{jython} or \emph{dotCMS}.

The average speedup is 24x,
ranging from a 4.7x speedup for \emph{batik}, one of the smallest benchmarks, to 52.68x for \emph{alfresco}.






\subsubsection{Completeness}
To evaluate the impact of \ourtechnique{} in analysis completeness, we first use the most straightforward coverage metric: the percentage of ``app reachable methods'' (i.e., methods deemed reachable in code that is part of the application or its immediate libraries and not in standard libraries) over all concrete methods in the application. The average reachability of app methods for the 2-obj analysis is \emph{64.50\%} without taking into account \emph{alfresco}, \emph{dotCMS}, and \emph{jython}, as these benchmarks timed out with a limit of 24 hours. Meanwhile, the 2-obj \ourtechnique{} analysis yielded \emph{61.30\%} app method reachability over the same benchmarks.

\begin{figure}[tb!]
  \includegraphics[scale=0.6, clip=true, trim=20 10 2 10]{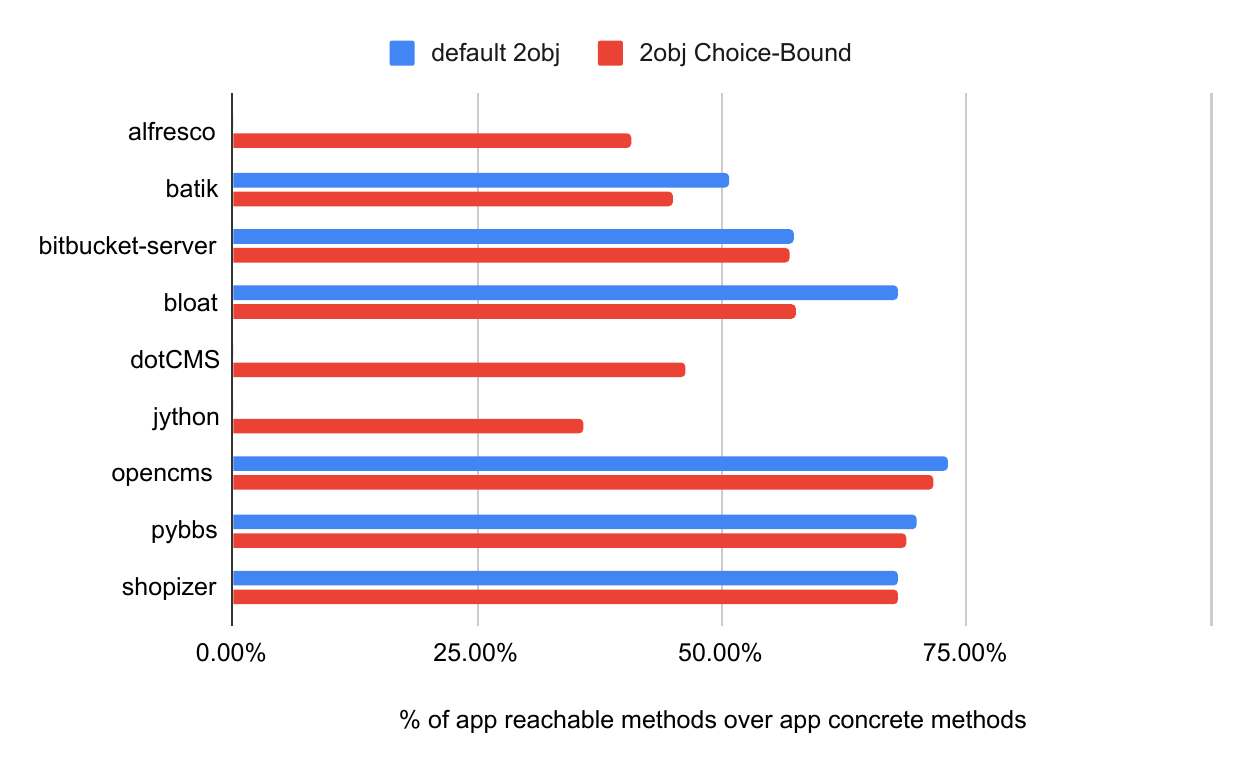}
  \caption{App methods reachability for the default 2-obj analysis and 2-obj with \ourtechnique{}.}
\label{fig:reachable}
\end{figure}

Figure~\ref{fig:reachable} plots the percentage of reachable methods in the application (plus the---non-system---libraries it is packaged with).
The coverage loss, in absolute difference, breaks down to:
\begin{itemize}
\item \emph{batik}: A loss of 5.63\% in app method coverage. 
\item \emph{bitbucket-server}: A loss of 0.41\% in app method coverage.
\item \emph{bloat}: A loss of 10.45\% in app method coverage.
\item \emph{opencms}: A loss of 1.62\% in app method coverage.
\item \emph{pybbs}: A loss of 0.99\% in app method coverage.
\item \emph{shopizer}: A loss of 0.05\% in app method coverage.
\end{itemize}

In most cases, the loss of app method coverage is negligible, especially in the case of web applications.

Consider that all numbers are obtained with a uniform parameterization in the design space of \ourtechnique{}, namely
\choicedomain{VarPointsTo}{var,ctx}{101}{hobj,hctx},
as discussed in Section~\ref{sec:applications}.
If one adapts the parameters per benchmark or per rough benchmark characteristics (e.g., large web applications vs. DaCapo
benchmarks), even better coverage and speedup should be possible. However,
the goal of our evaluation is to demonstrate that one can get very good results with little-to-no tuning,
for a globally uniform choice of parameters.

For a second, more targeted metric of analysis completeness, we consider one of the main reports
of \textsc{Doop}: the number of reachable casts that may potentially fail (in application-level code).
Since we have already quantified reachability (i.e., percentage of application methods that are covered by the analysis), we now
focus on the subset of methods 
that are reachable for both \ourtechnique{} and the 2-object-sensitive analysis, for each benchmark.
On the six benchmarks for which the 2-object-sensitive analysis terminates, we evaluate the loss of completeness as the number of application casts that may fail, where these casts are discovered by the 2-object-sensitive analysis but not by the \ourtechnique{} analysis.


\begin{figure}[tb!]
  \begin{tabular}{lp{4cm}p{2.5cm}p{2.5cm}}\toprule
  \textbf{Benchmark} &\textbf{app may fail casts for 2obj$\cap$\ourtechnique{} reachable methods} &\textbf{app \ourtechnique{} may fail casts} &\textbf{Completeness Loss} \\\midrule
  alfresco &N/A &3,499 &0.00\% \\
  batik &780 &726 &6.92\% \\
  bitbucket-server &121 &121 &0.00\% \\
  bloat &762 &748 &1.84\% \\
  dotCMS &N/A &4,496 &0.00\% \\
  jython &N/A &686 &0.00\% \\
  opencms &1,718 &1,681 &2.15\% \\
  pybbs &53 &53 &0.00\% \\
  shopizer &262 &252 &3.82\% \\  
  \bottomrule
  \end{tabular}
\caption[]{App May Fail Casts and Completeness Loss.}
\label{fig:appmayfailcaststable}
\end{figure}


Figure~\ref{fig:appmayfailcaststable} tabulates the results of the comparison.
As can be seen, the completeness loss identified in this case ranges between 0\% for \emph{pybbs} and 6.92\% for \emph{batik}.
Considering that a \textsc{Doop} analysis is unsound by default, the extra loss of identified \emph{application casts that may potentiall fail} 
is almost negligible between the two analyses.

Overall, the experiment with \textsc{Doop} validates that \ourtechnique{} offers the analysis designer a powerful tool in tuning scalability vs.
completeness. Although \ourtechnique{} will incur some completeness loss, this is typically minimal compared to the gains, with potential speedups of over an
order of magnitude, along with multiple cases where a normally non-terminating analysis becomes a very realistic sub-hour task.

\subsection{Symbolic Value-Flow Analysis}

We applied \ourtechnique{} to the symbolic value-flow (\emph{symvalic}) analysis framework~\cite{symvalic} for Ethereum VM smart contracts.
At first glance, this is a surprising domain of application, since the setting of smart contracts has fewer
scalability problems (compared to large Java applications):
smart contracts are small in size, only up to 24KB each in binary form.
However, the symvalic analysis itself is very precise (path-sensitive, with symbolic evaluation), therefore it does occasionally fail to scale.
In a sense, this analysis serves as validation of the universality of \ourtechnique{}, in diverse settings.
The framework itself is quite large, with some-3,000 Datalog rules, so it certainly
serves to validate the ease of application of the technique.

Notably, since the analysis is used to find several tens of real-world vulnerabilities, we can compare completeness a lot better,
with in-depth metrics instead of just coverage. (Although coverage remains a reliable indicator.)

The input dataset consists of all smart contracts on the Ethereum blockchain that are:
\begin{itemize}
\item deployed between Ethereum block number 21M (produced on Oct. 19, 2024) and 21.09M (produced on Nov. 1, 2024);
\item at least 15KB in size, to eliminate very small ``proxy'' contracts and other trivial contracts;
\item deduplicated by ``normalized bytecode'', i.e., if two contracts have the same code modulo configuration constants, only one is kept.
\end{itemize}

This yields 754 unique contract codebases. The Gigahorse decompiler~\cite{gigahorse}, underlying the analysis, fails to decompile
10 of the contracts (at least with its default settings). The remaining 744 comprise the input dataset.

We use a machine with two Intel Xeon E5-2687W v4 3.00GHz CPUs and 512 GB of RAM (each with 12 cores x 2 hardware threads).
We analyze 24 contracts at a time, each by a single thread. We set a timeout of 1500sec and a maximum RAM consumption of 50GB. (Most
analyses take a lot less memory than that---typically under 5GB---so no global RAM pressure arises throughout our benchmarking, even when some individual contract analyses reach
the limit of 50GB.)

\subsubsection{Performance}

The table in Figure~\ref{fig:symvalic-all} shows a summary of the performance of the symvalic analysis (core analysis + clients)
with and without \ourtechnique{} over all contracts in the input dataset. As can be seen, the use of
\ourtechnique{} virtually eliminates instances of out-of-memory and out-of-time execution, dropping the total to just 7 (out
of the 744 contracts) instead of 57 for the original analysis. In terms of average time, \ourtechnique{} results
in a 1.77x speedup. This is diluted by the large number of contracts with no scalability issues.

\begin{figure}
\begin{tabular}{|l|c|c|}
  \hline
     & \ourtechnique{} & Default symvalic analysis \\
  \hline
  out of time (1500sec) & 6 & 16 \\
  out of memory (50GB) & 1 & 41 \\
  \hline
  average analysis time & 98.1sec & 173.8sec \\
  \hline
\end{tabular}
\caption[]{Cumulative failures to analyze and analysis time for symvalic analysis, over 744 Ethereum smart contracts, with and without \ourtechnique{}.}
  \label{fig:symvalic-all}
\end{figure}

If we focus our attention to the smart contracts that run into problems with the default analysis
(i.e., the 57 contracts with either out-of-memory errors or 1500sec timeouts), \ourtechnique{}
exhibits an average execution time of 569.0sec. (It times out for 6 of these contracts and runs out of memory for 1, as shown
in the table.) The lower bound for the average speedup over this set is 2.53x.
(It is a lower bound both because executions that time out at 1500sec would take a lot longer if allowed to
complete, but also because we conservatively charge executions that run out of memory only the time it took them to reach the out-of-memory error and not
the full 1500sec.)

\subsubsection{Completeness}

We evaluate completeness in several different ways. The most straightforward is the overall analysis coverage
(in terms of block coverage and variable coverage, i.e., how many low-level program variables have values) for
all smart contracts in the input dataset.

\begin{figure}
\begin{tabular}{|l|c|c|c|}
  \hline
  & \ourtechnique{}  & \ourtechnique{}  & Default symvalic analysis \\
  & over all contracts & over common contracts & \\ 
  \hline
  block coverage & 92.9\% & 93.4\% & 93.5\%  \\
  variable coverage & 91.9\% & 92.5\% & 92.6\% \\
  \hline
\end{tabular}
\caption[]{Analysis coverage metrics over all successfully-analyzed smart contracts and over
  contracts analyzed successfully by both analyses.}
  \label{fig:symvalic-coverage-all}
\end{figure}

Figure~\ref{fig:symvalic-coverage-all} shows the standard coverage metrics, per analysis.
The first \ourtechnique{} column shows results for all successfully analyzed contracts (737 in total).
The second \ourtechnique{} column and the default analysis column show results for the 687 contracts analyzed
successfully by the default analysis (as well as by the \ourtechnique{} analysis).
As can be seen, \ourtechnique{} maintains high coverage over all contracts (e.g., 92.9\% block coverage vs. 93.5\% for the
original analysis) and, if we compare over the same set of contracts, there is virtually no loss of coverage (93.4\% vs. 93.5\% for block
coverage).
  


For a more in-depth comparison of completeness, we consider the vulnerability warnings that the analysis
issues, over the contracts analyzed by both configurations. Figure~\ref{fig:symvalic-bugs-common} tabulates the warnings issued with the two highest confidence
levels.
(The results for other confidence levels are very analogous, but larger in volume.)
As seen in the table, the report instances are nearly-identical, with very minimal completeness loss.

\begin{figure}
\begin{tabular}{|l|c|c|}
  \hline
  Warning type   & \ourtechnique{} & Default symvalic\\
  & &  analysis \\
  \hline
HIGH: Call to Tainted Function & 16 & 19 \\
HIGH: Chainlink data feed may provide stale answers& 52 & 52 \\
HIGH: DoS Call can cause failure & 8 & 8 \\
HIGH: FlashLoan unchecked callback & 4 & 4    \\
HIGH: Guard can be overwritten & 39 & 39   \\
HIGH: Inconsistent Reentrancy guards & 2128 & 2130 \\
HIGH: Merkle node can be used as leaf & 16 & 16   \\
HIGH: Rare tainted money-sensitive var in external call & 73 & 74   \\
HIGH: Reentrancy & 724 & 734  \\
HIGH: SSTORE to tainted address & 12 & 12   \\
HIGH: Stale value in storage & 20 & 20   \\
HIGH: Suspicious decimal arithmetic & 8 & 2    \\
HIGH: Swap publicly reachable & 533 & 533  \\
HIGH: Tainted Ownership Guard & 18 & 20   \\
HIGH: Tainted delegatecall & 35 & 35   \\
HIGH: Tainted money-sensitive var in external call & 1065 & 1081 \\
HIGH: Twin calls & 41 & 44   \\
HIGH: Unchecked Low-Level Call & 34 & 34   \\
HIGH: Uniswap price manipulation potential & 27 & 27   \\
HIGH: Uniswap tainted token & 2 & 2    \\
HIGH: Unrestricted approve proxy & 200 & 202  \\
HIGH: Unrestricted transfer proxy & 595 & 614  \\
HIGH: Unrestricted transferFrom Proxy & 7 &  7    \\
HIGH: this.call() & 47 & 47 \\
\hline
HIGHEST: Call to Tainted Function & 10 & 10 \\
HIGHEST: Inconsistent Reentrancy guards & 1983 & 1989 \\
HIGHEST: Rare tainted money-sensitive var in external call & 56 & 56 \\
HIGHEST: Reentrancy & 312 & 317 \\
HIGHEST: Stale value in storage & 12 & 12 \\
HIGHEST: Unrestricted transferFrom Proxy & 5 & 5 \\
\hline
\end{tabular}
\caption[]{Analysis high- and highest-confidence warnings over all successfully-analyzed smart contracts by both analyses: 687 total contracts.}
  \label{fig:symvalic-bugs-common}
\end{figure}

An interesting exception/observation concerns the results for
the \emph{Suspicious decimal arithmetic} warning category, where the \ourtechnique{} analysis yields more reports than the original. The reason
is that this client analysis uses negation over the main bounded predicates: the results of an arithmetic operation are not by themselves suspicious,
they are suspicious if \emph{no other} checks over the arguments exist in the code. Therefore, incomplete results in the core analysis predicates
yield \emph{more} final reports. Additionally, the program points that this analysis client checks are among the most central, heavily used in the code:
there are ``safe arithmetic'' routines in smart contracts, and these are called from multiple code points in the same smart contract. (E.g.,
every time there is a division operation in the source code, there is a call to a \sv{safeDiv} routine in the bytecode, which checks that the
division is not by zero.) Therefore this analysis client magnifies any incompleteness of the core symvalic analysis: the program sites that
are examined are the ones most likely to have lost information due to bounding of combinations. Still, the number of overall reports (8) is very
low compared to the total number of programs analyzed (687). This remains a useful analysis client, although it stands out as a ``worst case''
example of incompleteness of \ourtechnique{}.

In all other client analyses, the results are very close in number, with differences at the noise level.
This confirms that
\ourtechnique{} largely maintains the completeness of the original analysis when the original analysis scales well.

\newpage
\section{Related Work}

\paragraph{Program analysis optimization.}
There are numerous recent publications in making program analysis faster, with implementations purely in Datalog-based analyses.
All of these techniques (e.g., \cite{DBLP:journals/pacmpl/MaY0M0023,ZipperE,UnityRelay,Scaler,Hybrid,SelectiveImpactPre,ReturnofCFA,elephant} but also many more)
improve the core analysis logic, in an effort to avoid unscalability for specific cases. This is a great endeavor and very significant, since improving the
core analysis does more than avoid scalability issues: it often improves precision, i.e., yields a fundamentally better analysis, performance notwithstanding.
However, as means to improve \emph{performance}, such techniques still fall short of the goal: they neither provide a transparent, universal optimization,
nor address the problem in all instances.

Generally, \ourtechnique{} is orthogonal to any algorithmic improvements
in the analysis itself and its benefits can contribute to any other improvements. For instance, our evaluation of \ourtechnique{} was over
(a superset of) the enterprise application benchmarks that the authors of the JackEE analysis~\cite{elephant} assembled. Yet, unlike JackEE, \ourtechnique{}
achieves significant scalability benefits without having any special logic for Java enterprise applications or for custom versions of Java library
classes. Combining the two approaches can be done transparently and may well lead to ever further improvements. (This and other combination experiments
can be fruitful future work.)

\paragraph{Datalog with lattices.}
In terms of general, cross-cutting techniques, a modern trend that indirectly helps with scalability is to short-circuit an analysis by employing a \emph{lattice}
domain~\cite{DBLP:conf/pldi/SzaboEB21,DBLP:conf/pldi/MadsenYL16,DBLP:journals/pacmpl/SzaboBEV18}.  By introducing an ordering over the domain of abstract values, lattices allow the analysis to merge specific, fine-grained values into coarser ones, such as replacing a set of values with a top element ($\top$) when the set grows too large~\cite{DBLP:journals/pacmpl/SzaboBEV18,DBLP:conf/issta/MadsenL18,DBLP:conf/pldi/SzaboEB21}. This technique can short-circuit computations and prevent the explosion of intermediate results.

In practice, this approach does not adequately address the scalability challenges of high-performance Datalog analyses. First, porting an analysis from
a domain of explicit, enumerated values to a lattice domain may be non-trivial. There are language features that greatly automate the process. A prominent
representative is \souffle{}'s  \emph{subsumption}. Per the documentation\footnote{https://souffle-lang.github.io/subsumption} ``\emph{[s]ubsumption permits to delete more specific tuples by more general tuples. A programmer can express this by declaring a partial-order in the form of a subsumptive clause.}''

However, the implementation of subsumption incurs significant overheads. It requires checking the conditions for short-circuiting and managing the lattice operations.
At a high-level, every single tuple ever produced (for a relation that employs subsumption) needs to be checked to infer whether it is extraneous and should be
ignored in favor of a more general value. Specifically, in the case of \souffle{},
this necessitates the use of updatable data structures (such as BTreeDelete\footnote{\url{https://github.com/souffle-lang/souffle/blob/master/src/include/souffle/datastructure/BTreeDelete.h}}). The result is degraded performance due to the evaluation overhead, but possibly also increased complexity in memory management.

It is telling that, while experimental engines such as Flix~\cite{DBLP:conf/pldi/MadsenYL16,DBLP:conf/issta/MadsenL18} support lattices and subsumption, and \souffle{} offers a subsumption construct, these features have not been widely adopted in major third-party research or industrial analysis artifacts. Our own experiments with \souffle{}'s subsumption resulted in significant slowdowns, rather than speedups, for complex analyses. This limitation underscores the need for alternative techniques, such as our choice-based combination pruning, which can provide scalability without incurring the overhead associated with lattice operations.

\paragraph{Saturation-based techniques.}
The idea of stopping analysis at a certain threshold is fairly straightforward and used before in points-to analysis. Recent work by Wimmer et al.~\cite{saturation}
uses the evocative name ``\emph{saturation}''. The analysis stops enlarging the points-to set of a variable once it reaches a certain threshold in number of values.
\ourtechnique{} has significant differences: it is applied to a very different setting (declarative analysis instead of imperative, context-sensitive analyses instead
of context-insensitive that tries to become as fast as type-based); is transparent and virtually automatic instead of requiring intrusive changes; and
introduces a general framework for considering the design parameters, especially for context-sensitive analyses.

\souffle{} Datalog already supports a primitive form of saturation by means of the \sv{limitsize} construct, which stops evaluation of a relation when its
size (in tuples) reaches a threshold. The \sv{limitsize} construct is too crude for applications such as those of \ourtechnique{}, however. It limits
the whole relation and not combinations of its columns. As a result, it is very hard to invent appropriate \sv{limitsize} values for benefit without sacrificing analysis
completeness, especially if these values are not selected per-analyzed-program. Notably, the problem cannot be solved by just projecting out some columns
of the relation and applying \sv{limitsize} to the result: this would bound the projected relation, but not the original. \ourtechnique{} has the important
benefit of applying directly to the bounded relation itself, at the core of its evaluation.

\paragraph{Demand-driven and incremental analyses.}
Demand-driven program analyses (e.g., \cite{pldi/HeintzeT01,popl/ZhengR08,DBLP:conf/ecoop/SpathDAB16,Xu17,Sui16,oopsla/SridharanGSB05,mattmight:Sridharan:2006:Refinement,Shang12,Yan11}) compute information only for program points of interest
for a specific query. This line of work is related to \ourtechnique{} in the sense that a demand-driven analysis benefits from only needing \emph{part} of the results that
an analysis is capable of producing. However, sacrificing some completeness for performance is distinctly different from only needing some of the results to begin with.
\ourtechnique{} is aimed at analyses that do in principle make use of most of the potential results, and does not require fundamental changes to the analysis itself.

Similarly, incremental analyses (e.g., \cite{DBLP:journals/pacmpl/SzaboBEV18,DBLP:conf/pldi/SzaboEB21}) can avoid re-computing results
if these are guaranteed to not have changed, upon a small change of the input facts. Again, the intuition is similar: a part of the analysis inferences are
sufficient for computing the desired result. However, the setting of \ourtechnique{} is one of full analysis evaluation, making no assumptions on having small input changes.

\paragraph{Non-deterministic choice in databases and in logic programming.}
In Prolog, non-deterministic choice is inherent to the language's operational semantics. Prolog explores multiple execution paths through backtracking, allowing for the representation of non-deterministic computations naturally.

The integration of Prolog and relational database technology (used by modern high-performance Datalog implementations) was a longstanding goal for both the logic programming
and the database communities~\cite{Kun82}. Early attempts focused on straightforwardly applying a Prolog language processor to a relational database system~\cite{Kun82,Venken84,Berghel85}. However, these efforts faced challenges due to the mismatch between the computational models of the two systems. Prolog employs a tuple-at-a-time model of computation~\cite{Tsur85}, while relational databases operate on a set-at-a-time basis.

Datalog, as a purely declarative logic programming language, resolves the dilemma in favor of set-at-a-time computation.
This means that non-determinism does not come ``for free'' in the language. To introduce controlled non-determinism into Datalog, Giannotti et al.~\cite{NonDet} proposed a choice construct, allowing the expression of queries that require selecting arbitrary elements from a set. This construct enables the definition of functional dependencies within relations by enforcing a global constraint that, for a given key, only one tuple is selected. It is useful for tasks like breaking symmetry or selecting representative elements in data processing, such as in the implementation of greedy algorithms. \souffle{}~\cite{choice-domain}, as a modern Datalog engine, has incorporated the choice construct natively, building upon these foundational ideas. \souffle{}'s choice implementation is subsequently extended and leveraged in our approach to scale program analysis.

Other logic programming languages have explored similar concepts. IDLOG~\cite{Shen90} extends Datalog with a more sophisticated choice construct that enables sampling queries. For example, the following IDLOG program defines the sampling query: ``find an arbitrary set of employee samples which contains exactly two employees from each department.''

\begin{small}
\begin{verbatim} select_two_emp(Name) <- emp[2](Name, Dept, N), N < 2 \end{verbatim}
\end{small}

In this example, the notation \sv{emp[2]} specifies that, for each department, exactly two employees are selected based on the tuple identifier (tid). The condition \sv{N < 2} ensures that only the first two tuples (with tid 0 or 1) are considered in the computation of the relation \sv{select\_two\_emp}. This mechanism is similar to the SQL:2003 window function with a condition on \sv{ROW\_NUMBER()}~\cite{SQL2003}:

\begin{small}
\begin{verbatim}
SELECT name FROM (
  SELECT name, ROW_NUMBER() OVER (PARTITION BY dept) AS N FROM emp
) WHERE N < 3
\end{verbatim}
\end{small}

While such sampling queries or window functions can be useful for limiting results in non-recursive queries, they cannot be employed within a recursive stratum of a Datalog program (due to recursion through aggregation). They operate on already-computed relations and do not prevent the previous computation from generating all possible tuples. In contrast, \ourtechnique{} can be applied within the recursive evaluation itself, effectively preventing the computation of unnecessary tuples and thus enhancing scalability.

Relational databases provide features like \sv{LIMIT} or \sv{TOP} to restrict the number of tuples returned by a query. However, these mechanisms are crude and unwieldy when applied to recursive computations or complex analyses. They indiscriminately cut off the result set without considering the semantic importance of the data, leading to incomplete or inconsistent results. Moreover, they cannot be easily used to prevent the generation of large intermediate results during recursive evaluation since limits would be easily hit at early iterations of the semi-naive evaluation.

In summary, previous work has explored non-deterministic choice and techniques for limiting the size of computed relations, but these approaches either do not integrate seamlessly with recursive Datalog computations or introduce performance overheads that negate their benefits. Our approach leverages the choice construct in a novel way to bound the evaluation of predicates adaptively, providing a simple and effective solution to the scalability problem in Datalog-based program analyses.

\section{Conclusions} 

We presented \ourtechnique{}: a technique that offers
a powerful mechanism for tuning scalability vs. completeness in declarative computations.
\ourtechnique{} leverages an efficient, simple mechanism of non-deteministic choice in modern Datalog engines.
The idea is to implement multiplicity dependencies over existing relations: enforce that the same 
combination of pre-selected variables/columns can arise only up to a certain number of times. A large design
space arises from this simple principle, offering expressiveness and flexibility.

Applied to program analysis algorithms, which can unpredictably fail to scale, \ourtechnique{} has significant
value: previously unscalable analyses can now become entirely realistic, at the expense of a small loss in
completeness. (More non-determinism is also introduced, but non-determinism, from several sources, is inevitable
in all analyses we have encountered.) We apply \ourtechnique{} to pre-existing, large Datalog program analysis frameworks, such as
\textsc{Doop} and its ``ideal'' analysis, 2-object-sensitive+heap. In subject programs also examined in
past literature, for which the default analysis had difficulties scaling, \ourtechnique{} achieves speedups
typically well over 10x. Analyses that would not terminate in over 24 or 48 hours now run in sub-hour time.
The result is a powerful tool in the hands of an analysis designer, permitting customization and obtaining
results even for highly-complex analyses and large inputs.

\bibliographystyle{ACM-Reference-Format}
\bibliography{bibliography,references,tools,bib/ptranalysis,bib/proceedings}

\end{document}